\documentclass[letterpaper]{JHEP3}

\usepackage{epsfig} %For EPS graphics

\newcommand{\pd}{\ensuremath{\partial}}
\newcommand{\Tr}{\ensuremath{\textrm{Tr}}}

\newcommand{\be}{\begin{equation}}
\newcommand{\ee}{\end{equation}}
\newcommand{\bea}{\begin{eqnarray}}
\newcommand{\eea}{\end{eqnarray}}

\title{Collective Field Description of Matrix Cosmologies}
\author{Morten Ernebjerg\thanks{morten@physics.harvard.edu}, Joanna L. Karczmarek\thanks{karczmar@fas.harvard.edu} and Joshua M. Lapan\thanks{lapan@fas.harvard.edu}\\ Jefferson Physical Laboratory,
Harvard University, Cambridge, MA, 02138} \abstract{We study the
Das-Jevicki collective field description of arbitrary classical
solutions in the $c=1$ matrix model, which are believed to
describe nontrivial spacetime backgrounds in 2d string theory. Our
analysis naturally includes the case of a Fermi droplet cosmology:
a finite size droplet of Fermi fluid, made up of a finite number
of eigenvalues. We analyze properties of the coordinates in which
the metric in the collective field theory is trivial, and comment
on the form of the interaction terms in these coordinates.}

\keywords{Matrix Models, Tachyon Condensation} 
\preprint{0405187}
\begin{document}

\section{Introduction}
\label{sec:intro}

The soluble string theory in 1+1 dimensions is a rich toy model for
the study of nonperturbative phenomena often not accessible to
analysis in higher dimensional theories. Of such phenomena, one
class are the processes with a nontrivial time evolution. Examples
include tachyon condensation, creation and evaporation of black
holes, and cosmological evolution.

An important step towards the study of time-dependent phenomena in
the $c=1$ matrix model for the 2d string was the description of
D0-brane decay, or open string tachyon condensation
\cite{Klebanov:2003km,McGreevy:2003ep}. The matrix model provides an
exact picture of the time evolution as the classical motion of a
single matrix eigenvalue; its predictions were compared to
worldsheet string analysis and found to agree.

Classical collective motions of the entire Fermi sea, as opposed to
a motion of a single eigenvalue, were described for example in
\cite{Minic:rk,Alexandrov:2002fh}.  These describe nontrivial time
dependent backgrounds for the 2d string theory and were interpreted
as closed string tachyon condensation in \cite{Karczmarek:2003pv}.
Another class of time-dependent solutions---droplets of large but
finite number of eigenvalues, corresponding to closed universe
cosmologies---was proposed in \cite{Karczmarek:2003pv}. Since  these
classical time-dependent solutions of the matrix model correspond to
large motions of the Fermi surface, small fluctuations about the
Fermi surface carry important information about propagation of
stringy spacetime fields. As we will review below, these small
fluctuations can be described in the Das-Jevicki collective field
approach by a 2d effective field theory, whose action generically
contains a nontrivial, time-varying metric.

A step toward understanding these time-dependent solutions was taken
by Alexandrov in \cite{Alexandrov:2003uh}, where coordinates were
found in which the metric was made trivial.  However, the method
presented there does not extend to compact Fermi droplets.  The main
purpose of this note is to extend the construction of Alexandrov
coordinates to arbitrary Fermi surfaces, including compact cases.

To this end, we study the Alexandrov coordinates in some detail. In
section \ref{sec:notation}, we briefly review the collective field
description of small fluctuations about a time-dependent Fermi
surface.  In section \ref{sec:existence}, we explicitly construct
the Alexandrov coordinates for an arbitrary solution. In section
\ref{sec:static}, we analyze a special class of backgrounds
(including some compact cases) for which the entire, interacting
action is static. The collective field action for these solutions is
shown to take a standard form with a time-independent coupling
constant. In section \ref{sec:droplet}, we construct the Alexandrov
coordinates for a finite collection of fermion eigenvalues: a
compact droplet cosmology.  We briefly discuss the possibility of
formulating a spacetime string theory interpretation of such a
configuration. Finally, in the Appendix, we analyze the interaction
term in the effective action and show by an explicit example that it
is not always possible to make it static.

\section{Notation and Alexandrov Coordinates}
\label{sec:notation}

In the double scaling limit, matrix quantum mechanics is defined
by the action \be S = {1 \over 2} \int dt ~\Tr \left(\dot M(t)^2 +
M(t)^2\right)~, \ee where $M$ is a Hermitian matrix whose size in
this limit is taken to infinity.  As is well known (for reviews,
see for example \cite{Polchinski:1994mb,Klebanov:1991qa}), upon
quantization the singlet sector of the matrix quantum mechanics is
described by an infinite number of free (noninteracting),
nonrelativistic fermions representing the eigenvalues of $M$. The
fermions inherit the same potential as the matrix $M$, and hence
the single variable Hamiltonian is \be\label{eq:H} H = {1 \over 2}
(p^2 - x^2)~. \ee Since the number of fermions is large, the
classical limit of the theory is that of an incompressible Fermi
liquid moving in phase space $(x,p)$ under the equations of motion
given by the Hamiltonian (\ref{eq:H}).  We will restrict our
analysis in this note to situations where the Fermi surface (the
boundary of the Fermi sea) can be given by its upper and lower
branch, which we will denote with $p_{\pm}(x,t)$, see figure
\ref{figure}. It is easy to show that $p_{\pm}(x,t)$ satisfy
\be\label{eq:EOMp}
\partial_t p_\pm + p_\pm \partial_x p_\pm = x~.
\ee \FIGURE{\epsfig{file=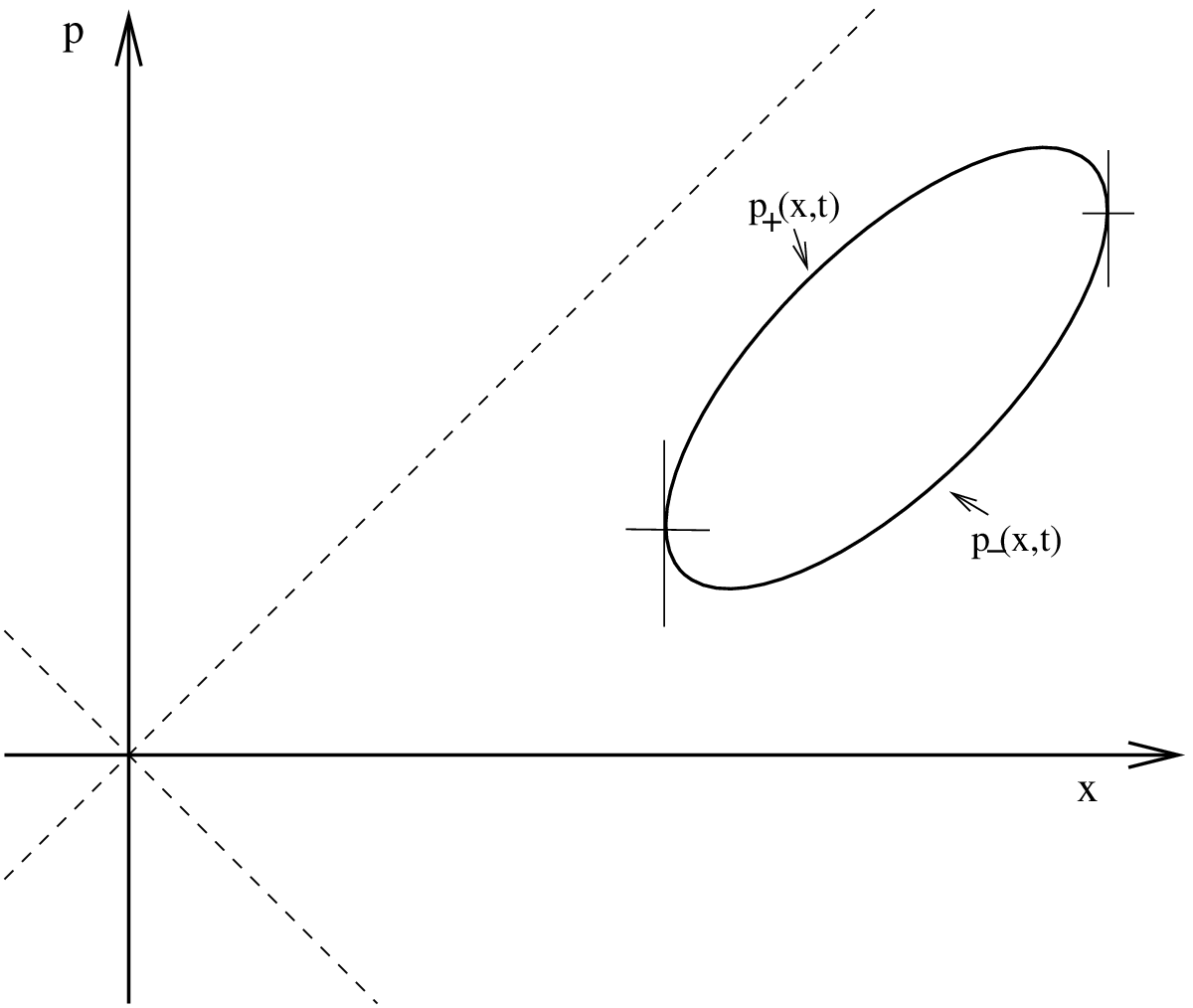}\caption{A compact Fermi
surface in phase space. The upper and lower branches of the
surface are labelled, and vertical points where they meet (and the
collective theory becomes strongly coupled) are
marked.}\label{figure}}
One way to directly connect the classical
limit of the matrix quantum mechanics with the collective
description of fermion motion is via a procedure developed by Das
and Jevicki \cite{Das:1990ka}. Define a field $\varphi (x,t)$ by
\be \varphi (x,t) = \frac{1}{\pi}\Tr[\delta (x-M(t))]~ \ee so that
$\varphi (x,t)$ is the density of eigenvalues at point $x$ and
time $t$. In the fermion description, we have the relation \be
\label{eq:phi} \varphi =  {p_+ - p_- \over 2}~. \ee The action for
the collective field is \cite{Das:1990ka} \be S = \int\!
\frac{dt\,dx}{2\pi} \left\{ \frac{Z^2}{\varphi} -
\frac{1}{3}\varphi^3 + (x^2 - 2\mu)\varphi \right\} ~, \ee where $
Z = \int dx \partial_t \varphi $, so the equation of motion is \be
\label{eq:phi_eoms}
\partial_t \left( \frac{Z}{\varphi} \right)
-
\frac{Z}{\varphi}\partial_x \left( \frac{Z}{\varphi} \right)
=
\varphi \partial_x\varphi - x~. \ee
Furthermore, we have the relation \cite{Alexandrov:2003ut}
\be \label{eq:Z_over_phi}
\frac{Z}{\varphi}=-\frac{p_+ + p_-}{2}~,
\ee
which allows us to verify that (\ref{eq:phi_eoms}) is
consistent with (\ref{eq:EOMp}).

We want to consider a fixed solution $\varphi_0(x,t)$ of
(\ref{eq:phi_eoms}) and study the effective action for small
fluctuations about this solution.  In the string theory dual to
the matrix model, this corresponds to studying the small
fluctuations about a string background given by the solution
$\varphi_0(x,t)$.  Let $\pd_x\eta(x,t)$ denote the small
fluctuations \be \varphi = \varphi_0 + \sqrt{\pi}\partial_x\eta
\ee and let $Z_0=\int dx \partial_t\varphi_0$.

Rewriting the action and grouping terms in powers of $\eta$ we find
(noticing that terms linear in $\eta$ vanish by the equations of motion)
\bea \label{eq:action1}
S & = & \int\! \frac{dt\,dx}{2\pi}  \left\{ \frac{(Z_0 + \sqrt{\pi}\partial_t\eta)^2}{\varphi_0 + \sqrt{\pi}\partial_x\eta} - \frac{1}{3}(\varphi_0+\sqrt{\pi}\partial_x\eta)^3 + (x^{2}-2\mu)(\varphi_0+\sqrt{\pi}\partial_x\eta)  \right\}  \nonumber \\
& = & S_{(0)}+S_{(2)}+S_{int} \eea where $S_{(0)}$ has no
$\eta$-dependence, \be\label{eq:S.quadratic} S_{(2)} = {1\over2} \int
{dt ~dx \over \varphi_0} \left\{ \left(\partial_t \eta - {Z_0\over
\varphi_0} \partial_x \eta\right)^{2} -\varphi_0^2 (\partial_x
\eta)^2 \right\}~, \ee and \bea S_{int}&=&\frac{1}{2}\int
\frac{dt~dx}{\varphi_0}\left\{-\frac{\sqrt{\pi}}{3}\varphi_0(\partial_{x}\eta)^3\right. \nonumber \\
& & \qquad \left.+\left(\partial_{t}\eta-\frac{Z_0}{\varphi_0}\partial_x\eta\right)^{2}\sum_{n=1}^\infty
(-\sqrt{\pi})^n \left(\frac{\partial_x\eta}{\varphi_0}\right)^n
\right\}~.
  \label{eq:interac}\eea In \cite{Alexandrov:2003uh}, it is proposed that
coordinates $(\tau,\sigma)$ exist in which $S_{(2)}$ takes a
standard form of a kinetic term for a field in a flat metric
\be\label{eq:S.standard} S_{(2)} = \int\! d\tau^+ d\tau^-\,
\partial_{\tau^-} \eta\,
\partial_{\tau^+} \eta~,
\ee where $\tau^\pm (x,t) = \tau \pm \sigma$ are the lightcone
coordinates. We shall refer to the coordinates ${(\tau,\sigma)}$
as the \emph{Alexandrov coordinates}. In \cite{Alexandrov:2003uh},
these coordinates were constructed from a specific form of the solution
$\varphi_0$.  In the next section, we prove (at least locally)
their existence for all $\varphi_0$.

\section{Alexandrov Coordinates -- Existence}
\label{sec:existence}

It is quite simple to show, using the equations of motion for the two
branches of the solution (\ref{eq:EOMp}),
that the action (\ref{eq:S.quadratic})
takes on the form in (\ref{eq:S.standard}) as long as the
coordinates $\tau^{\pm}$ satisfy
\be
\label{eq:tau.pm}
(\partial_t + p_\pm\partial_x) \tau^{\pm} = 0 ~.
\ee
Equation (\ref{eq:tau.pm}) can easily be solved (at least locally).
The exact form of the
solution depends on whether the slope of the solution $p_\pm$ is steeper
or shallower than 1.  The regions where $\alpha(x,t) \equiv
\partial_x p_\pm$ satisfies $|\alpha| > 1$ will be referred to as the
steep regions, and those were $|\alpha| < 1$ will be referred to as
the shallow regions.  In the steep regions, we have that \be
\label{eq:steep} \tau^\pm = t- \coth^{-1} \left(\partial_x p_\pm
\right)~, \ee and in the shallow regions we get \be
\label{eq:shallow} \tau^\pm = t-\tanh^{-1} \left(\partial_x p_\pm
\right)~. \ee The solution above is not unique---a conformal change
of coordinates does not change the form of the quadratic part of the
action (\ref{eq:S.standard}), so any change of coordinates of the
form \be \tau'^\pm = \tau'^\pm(\tau^\pm) \ee will provide another
solution to equation (\ref{eq:tau.pm}).  For example, the following
is also a good solution \be
\tau^\pm = {\tanh t -
\partial_x p_\pm  \over 1- \partial_x p_\pm \tanh t} \ee
as is \be
\tau^\pm = {\coth t - \partial_x p_\pm \over 1-
\partial_x p_\pm \coth t}~. \ee
Note that if $p_+$ and $p_-$ are flat on some overlapping region
($\partial_x p_{\pm}=0$), then these coordinates will be degenerate.
However, we can easily parameterize these flat regions in a
nondegenerate way so that the metric is still flat.

The solutions (\ref{eq:steep}) and (\ref{eq:shallow}) are valid
locally on steep and shallow coordinate patches respectively
(modulus the degenerate case mentioned above). To create a single
coordinate system, we can `glue together' the various steep,
shallow, and flat patches by using the freedom of conformal
coordinate changes. While our expressions guarantee the existence of
Alexandrov coordinates on each patch and while there are no obvious
obstacles to the `gluing' procedure, constructing the coordinates in
this way would be very cumbersome, even in cases where the resulting
coordinate systems are simple.  Instead, in all the examples given
in this paper, the Alexandrov coordinates are constructed by the
procedure given in \cite{Alexandrov:2003uh} (but see the comment at
the end of section \ref{sec:droplet}).

Another issue is that, as was shown in \cite{Alexandrov:2003uh}, the
resulting coordinates often have boundaries (this will also be seen
in section \ref{sec:droplet}).  The boundaries come in two
categories.  The first are timelike boundaries corresponding to the
end(s) of the Fermi sea; the boundary conditions on those can be
determined from the conservation of fermion number
\cite{Das:1990ka}. The second class of boundaries contains
boundaries which are either spacelike or timelike, do not have a
clear interpretation, and for which appropriate  boundary conditions
are not known.  We will return to the issues of boundaries in the
discussion in section \ref{sec:droplet}.

We will close this section with a simple example as an
illustration. Consider a moving hyperbolic Fermi surface given
parametrically by \cite{Karczmarek:2003pv} \bea
\label{eq:movingpara}
x &=& \sqrt{2\mu} \cosh{\sigma} + \lambda e^t \nonumber \\
p &=& \sqrt{2\mu} \sinh{\sigma} + \lambda e^t~. \eea In
this case, we have $\varphi_0 = \sqrt{(x- \lambda e^t)^2 - 2\mu} =
\sqrt{2\mu} \sinh \sigma$.  The Alexandrov coordinates are given
simply by $\sigma$ in the parametrization above and by $\tau = t$.
It is a simple matter to check that the action takes the form \bea
S&=&\label{eq:action.sinh} \int d\tau d\sigma \left\{ {1\over 2}
((\partial_\tau\eta)^2 - (\partial_\sigma \eta)^2) - { \sqrt{\pi}
\over 6 \varphi_0^2} (3(\partial_\tau\eta)^2(\partial_\sigma
\eta)+ (\partial_\sigma \eta)^3) \right. \nonumber \\ &
&\qquad\left. + \frac{(\partial_\tau \eta)^2}{2}\sum_{n=2}^{\infty}
 \left(-{\sqrt{\pi} (\partial_\sigma \eta) \over \varphi_0^2}\right)^n
\right\}~. \eea Note that the coupling diverges at the point
$\sigma=0$ which corresponds to the edge of the Fermi sea, and
that it does not depend on $\tau$.

\section{Alexandrov Coordinates -- Special Case}
\label{sec:static}

In this section, we study a class of solutions (of which an example
appeared at the end of the previous section) for which the
Alexandrov coordinates can be written as \be \label{eq:ansatz}
\sigma =\sigma(x,t)~, \qquad \tau=\tau(t)~. \ee We shall see that
this leads to a very restricted class of solutions, but a class
which includes both infinite and finite (compact) Fermi seas. Thus,
it encompasses the two generic types of dynamic solutions.

With the coordinate ansatz above, we have \bea dt\,dx &=& \frac{
d\tau\, d\sigma }{| \pd_{x} \sigma\, \pd_{t} \tau|}
\nonumber\\
\pd_{x} &=& \pd_{x} \sigma\,\pd_{\sigma} \nonumber\\
\pd_{t} &=& \pd_{t} \tau\,\pd_{\tau} + \pd_{t}\sigma\,\pd_{\sigma}~.
\eea Demanding that the kinetic term take the standard
flat form \be S_{(2)}=\int
d\tau\,d\sigma\left[\frac{1}{2}\left(\pd_{\tau}\eta\right)^{2}
-\frac{1}{2}\left(\pd_{\sigma}\eta\right)^{2}\right] \ee leads to
the requirements that \be \pd_{t}\sigma
=\frac{Z_{0}}{\varphi_{0}}\pd_{x}\sigma \qquad \textrm{and} \qquad
\left|\frac{\pd_{t}\tau}{\pd_{x}\sigma} \right| =\varphi_{0}~.
\label{eq:req}\ee

These constraints can be solved explicitly, provided that the
solution is only vertical at endpoints ($\varphi_0=0$). Since $\tau$
depends only on $t$, we find that $\partial_t \tau = (\partial_\tau
t)^{-1}$, $\partial_x \sigma = (\partial_\sigma x)^{-1}$,
$\partial_x \tau =
\partial_\sigma t = 0$,
\be
\partial_\tau x = -\frac{\partial_t \sigma}{(\partial_x \sigma )
(\partial_t \tau )}~, \qquad \textrm{and} \qquad \partial_t \sigma
= -\frac{\partial_\tau x}{(\partial_\sigma x )(\partial_\tau t )}~.
\ee
Using the first equation in (\ref{eq:req}) we find
\be
\partial_x Z_0  =
   - \frac{1}{\partial_\tau t}\left[\varphi_0 \partial_\tau \ln
(\partial_\sigma x) + \frac{\partial_\tau x}{\partial_\sigma x}
\partial_\sigma \varphi_0\right]~,
\ee
which is equal to
\be
\partial_x Z_0  =
\partial_t \varphi_0 =
\frac{1}{\partial_\tau t}\left[\partial_\tau \varphi_0 - \frac{\partial_\tau x}
{\partial_\sigma x}\partial_\sigma \varphi_0\right]~.
\ee
Comparing these two expressions, we obtain a
differential equation for $\varphi_0$
\be
\partial_\tau \ln (\varphi_0) = - \partial_\tau \ln (\partial_\sigma x)
\ee whose solution is clearly of the form $\varphi_0 = f(\sigma)^2
(\partial_\sigma x)^{-1}$.  This we can rewrite, using the second
equation in (\ref{eq:req}), as \be \varphi_0(\sigma, \tau) =
f(\sigma ) \sqrt{g(\tau )}~, \ee where $g(\tau )=(\partial_\tau
t)^{-1}$ and we assume $f(\sigma)>0$. We also have $\partial_x\sigma
=\sqrt{g}/f$.

Now we can use the equation of motion (\ref{eq:phi_eoms}) to find
the forms of $f$ and $g$.  Using (\ref{eq:req}), notice that
\be
\partial_\tau=(\partial_\tau t)\partial_t + (\partial_\tau x)\partial_x = (\partial_\tau t)\left( \partial_t - \frac{Z_0}{\varphi_0}\partial_x \right)~,
\ee
so the equation of motion (\ref{eq:phi_eoms}) implies
\be
\partial_\sigma \partial_\tau \left( \frac{Z}{\varphi} \right)
=
\partial_\sigma(\varphi \partial_x\varphi) -
{1\over \partial_x\sigma}~. \ee
Substituting the explicit form of $\varphi_0$ in terms of $f$ and $g$
into this equation, we obtain

%Combining this with (\ref{eq:phi_eoms}) yields \bea
%\partial_\sigma \partial_\tau \left( \frac{Z_0}{\varphi_0} \right)
%& = &  \frac{f}{4g^{3/2}}
%\left( 2g\partial_\tau^2 g - (\partial_\tau g)^2 \right)
%\nonumber \\
%& = & \frac{1}{g}\partial_\sigma \left( g^{3/2}\partial_\sigma f -
%x\right) = \sqrt{g}\partial_\sigma^2 f - \frac{f}{g^{3/2}}~, \eea
%so we find

\be 2g\partial_\tau^2 g - (\partial_\tau g)^2 + 4 -
4g^2\frac{\partial_\sigma^2 f}{f}=0~. \ee
Since $g$ only depends
on $\tau$, and $f$ only on $\sigma$,
we see that $\partial_\sigma^2 f = -\alpha f$ where
$\alpha$ is a constant.

Consider first the situation
when $\alpha$ is positive.  Then
\be f(\sigma )
= f_1 \sin (\sqrt{\alpha}(\sigma-\sigma_1) )~, \ee
where $f_1$ and $\sigma_1$ are real numbers.
To ensure that $\varphi_0\geq 0$, we must restrict
$f_1>0$ and $\sigma_1\leq\sigma\leq\sigma_1+\frac{\pi}{\sqrt{\alpha}}$.
Requiring $g$ to be real yields \be
\label{eq:g} g
(\tau ) = \frac{1}{\sqrt{\alpha}}\left[ \sqrt{c^2+1}\,
\cos(2\sqrt{\alpha}(\tau-\tau_1)) + c \right]~, \ee where $c$ and
$\tau_1$ are real constants of integration.
If $\alpha$ is negative and $|c|\geq1$, we have
\bea
f(\sigma )
&=& f_1 \sinh (\sqrt{|\alpha|}\sigma )+ f_2 \cosh (\sqrt{|\alpha|}\sigma )
~~{\rm{and}}\nonumber\\
g(\tau)&=& \frac{1}{\sqrt{|\alpha|}}\left[ \sqrt{c^2-1}\,
\cosh(2\sqrt{|\alpha|}(\tau-\tau_1)) + c \right]~,
\eea
while for $|c|<1$
\be
\label{eq:g1} g
(\tau ) = \frac{1}{\sqrt{|\alpha|}}\left[ \sqrt{1-c^2}\,
\sinh(2\sqrt{|\alpha|}(\tau-\tau_1)) + c \right]~.
\ee
Notice that positivity of $f(\sigma)$ restricts the choice of $f_{1}$ and $f_{2}$ while positivity of $g(\tau)$ in some of these cases restricts the range of $\tau$ to a finite or semi-infinite interval.

Let $F(\sigma ) = \int d\sigma f(\sigma )$ so that
$x(\sigma ,\tau )=(F(\sigma ) + k(\tau ))/\sqrt{g(\tau )}$ for some
function $k(\tau)$.   We can also show that $Z_0/\varphi_0$
is of the form
\be
\frac{Z_0}{\varphi_0} = h(\tau) -  {\partial_\tau
g \over 2\sqrt g} F ~.
\ee
The functions $h(\tau)$ and $k(\tau)$ can be computed using the
equation of motion.
Computing $p_\pm$ from (\ref{eq:phi}) and (\ref{eq:Z_over_phi}),
we get the following relationship

\be \alpha g^2\left(x-\frac{k(\tau)}{\sqrt{g(\tau)}}\right)^2 +
\left(p+x\frac{\partial_\tau g(\tau)}{2} + h(\tau) \right)^2 =
f_1^2 g(\tau) \ee which we recognize as an ellipse (a hyperbola)
if $\alpha$ is positive (negative).  Notice that, from equation
(\ref{eq:g}), the compact (elliptical) solutions correspond to a
finite range of $\tau$.

The interaction terms (\ref{eq:interac}) simplify under our
assumption to \be \label{eq:Sstatic} S_{int}=\int d\tau\,d\sigma
\left[ \frac{1}{6}\Lambda\!\left( \pd_{\sigma}\eta \right)^{3}
+\frac{1}{2}\left(\pd_{\tau}\eta\right)^{2}\sum_{n=1}^\infty
\Lambda^{n} \left(\pd_{\sigma}\eta\right) ^{n} \right]~, \ee where
the effective coupling constant is \be \Lambda =
-\frac{\sqrt{\pi}}{\varphi_{0}}\pd_{x}\sigma =
-\frac{\sqrt{\pi}}{f(\sigma)^{2}}~.\ee  So we find that the coupling
constant is time-independent for this class of solutions. We note
that the moving-hyperbola solution (\ref{eq:action.sinh}) falls into
this class.

As long as $|\Lambda\,\pd_{\sigma}\eta| <1$, we can sum the
series to get \be S_{int}=\int d\tau\,d\sigma \left[
\frac{1}{6}\Lambda\left( \pd_{\sigma}\eta \right)^{3}
+\frac{1}{2}\left(\pd_{\tau}\eta\right)^{2}\left(\frac{
\Lambda\pd_{\sigma}\eta}{1-\Lambda \pd_{\sigma}\eta}\right)
\right]~. \ee  The first interaction term diverges as
$\varphi_{0}\rightarrow 0$, which occurs when the width of the Fermi
sea goes to zero.  This corresponds to strong coupling at the tip of
the static hyperbolic Fermi surface. The second interaction term
diverges as $|\Lambda\,\pd_{\sigma}\eta| \rightarrow 1$. We have \be
\Lambda\,\pd_{\sigma}\eta =
-\frac{\sqrt{\pi}}{\varphi_{0}}\pd_{x}\eta =
\frac{\varphi_{0}-\varphi}{\varphi_{0}}~, \ee so the breakdown
happens when the excitations become comparable to the width of the
Fermi sea (as can also been seen directly from (\ref{eq:action1})).
In this case, the Fermi sea may pinch and split into two, so we
would not expect to be able to neglect interactions between the
upper and lower Fermi surface. Thus, the collective theory becomes
strongly coupled exactly in the places one would expect it to from
general considerations.

We have demonstrated that, under the restriction
(\ref{eq:ansatz}), the action takes a universal, static form
(\ref{eq:Sstatic}). The natural question to ask is whether such a
universal form of the action might exist for all solutions.  As a
partial answer to this question, in the Appendix we analyze
explicitly an example which does not fall into the class of
solutions studied in this section. We show that, even with the
freedom of conformal change of coordinates, it is not always
possible to make the interaction term static in Alexandrov
coordinates.

\section{Fermi Droplet Cosmology}
\label{sec:droplet}

In this last section, we construct an explicit example of the class
of solutions discussed above---a droplet solution in which only a
finite region of phase space is filled (so that the Fermi surface is
a closed curve).  Such solutions are believed to give rise to time
dependent backgrounds in the spacetime picture
\cite{Karczmarek:2003pv}, although no precise correspondence has
been found so far.

In the simplest case, the Fermi surface is a circle in phase space
with radius $R$ and center $(p,x)=(0,x_0)$ at time $t=0$. Notice
that we must demand $x_{0}>\sqrt{2}R$ in order for the surface not
to cross the diagonals $p=\pm x$ (otherwise, some of the fermions
will spill over the potential barrier as the droplet bounces off
it).

It is not difficult to write down the evolution of this
Fermi surface
\be
e^{-2t}(x+p-x_{0}e^{t})^{2}+e^{2t}(x-p-x_{0}e^{-t})^{2}=2R^{2}~.
\ee  Solving for $p$ we find \be \varphi_{0} =
\frac{\sqrt{R^2\cosh 2t-(x-x_0\cosh t)^2}}{\cosh 2t}~. \ee A
sensible $\sigma$-coordinate is an angle parameterizing the upper
surface, running from $0$ to $\pi$ between the points where
$\varphi_{0}=0$. These are given by \be x=x_{0} \cosh t\pm
R\sqrt{\cosh 2t}~, \ee so the simplest guess for an Alexandrov
coordinate (which we call $\theta$ to stress its angular nature)
is such that \be x=x_{0} \cosh t-R\cos \theta \sqrt{\cosh 2t}~.
\ee Using the second condition in (\ref{eq:req}), we find \be
\pd_{t}\tau=\frac{1}{\cosh 2t}~, \ee which gives \be \tau =
\tan^{-1}(\tanh t)~. \ee Thus, $\tau$ runs over the finite range
$-\pi/4\leq \tau \leq \pi/4$.  In these new coordinates, we find \be
x=\frac{1}{\sqrt{\cos 2\tau}}(x_0\cos\tau-R\cos\theta)~, \qquad
\varphi_{0} = R\sqrt{\cos 2\tau}\sin\theta~. \ee It can be checked
that these coordinates do fulfill the first condition
in (\ref{eq:req}) as well.

We see that \be \Lambda
=-\frac{\sqrt{\pi}}{\varphi_{0}}\pd_{x}\theta
=-\frac{\sqrt{\pi}}{R^2\sin^2\theta}~, \ee and \be
g(\tau)=\cos 2\tau~, \qquad f(\theta)=R \sin\theta~,\ee
and the action (\ref{eq:action1}) simplifies to
\bea
S &=& \int d\tau~d\theta~ \left\{  \frac{1}{2}[(\partial_\tau\eta)^2-(\partial_\theta\eta)^2] - \frac{\sqrt{\pi}}{6 R^2\sin^2\theta}(\partial_\theta\eta)^3   \right.  \nonumber \\
& & \left. + \frac{1}{2}(\partial_\tau\eta)^2\sum_{n=1}^\infty
\left(-\frac{\sqrt{\pi}}{R^2\sin^2\theta}\partial_\theta\eta\right)
^{\!n}  \right\}~. \eea
As anticipated, the theory is strongly coupled at the endpoints of
the droplet where $\varphi_{0}\rightarrow 0$.  Note that the
coordinates are smooth across the steep/shallow divide\footnote{It
is possible to explicitly reach the $\theta$-coordinate from the
generally applicable forms (\ref{eq:steep}), (\ref{eq:shallow}) by using
appropriate conformal transformations on each patch, but the computation is
complicated.}.

As an aside,  consider a modification to the droplet discussed
above. At time $t=0$, replace the regions $\pi/4 < \theta < 3\pi/4$
and $5\pi/4<\theta<7\pi/4$ by straight lines so that the droplet
takes the form of a rectangle with semi-circular ends. A
straightforward computation leads to the conclusion that one can
find global coordinates which yield a flat kinetic term in the
action. As one might expect, time is still compact as it was in the
elliptical case, indicating that the compactness is not merely an
accident occurring only for this particular shape.

While the droplets are amusing objects in matrix theory, it would
be more interesting and satisfying  if they had a clear spacetime
interpretation in string theory. The collective field description,
which we have constructed here, suggests that they have an
interpretation as some closed string backgrounds.  The massless
scalar fluctuations should correspond to some string field, the
analog of the tachyon in $c=1$ Liouville string. The strongly
coupled regions at each end of the droplet should correspond to
`tachyon walls'---strongly coupled regions of large tachyon VEV.
If such a closed string, worldsheet description could be found, it
would provide an example of an open-closed string, finite $N$
duality between a time-dependent finite universe and the matrix
quantum mechanics of the $D0$ branes making up the droplet.

Unfortunately, it is not clear how to construct such a spacetime
interpretation.  The natural time $\tau$ is compact, corresponding
to the fact that in the fermion time, $t$, fluctuations of a compact
Fermi surface become frozen in the past and future.  Also, examining
the interaction term, we notice that the coupling $\Lambda$ is
bounded from below so that the theory does not approach a free
theory in any region (though the coupling can be made arbitrarily
small by taking $R$ large).  This makes it unlikely that it will be
possible to define an S-matrix. In addition, in the standard $c=1$
story, the matrix-to-spacetime dictionary is complicated by the
presence of leg pole factors, additional phases needed to match the
matrix model S-matrix to its string worldsheet counterpart.
Supposedly such a complication would appear for the droplet
cosmologies as well, but there is no obvious candidate for what it
might be. Therefore, it seems unlikely that a spacetime analysis of
tachyon scattering can be carried out as has been done in the case
of the standard, static Fermi sea as well as the moving hyperbola
solution (\ref{eq:movingpara}) \cite{Karczmarek:2004ph, Das:2004hw}.

Another way to view the complication introduced by the finite extent
of the time $\tau$ is that in Alexandrov coordinates there appear
boundaries (in this case, spacelike boundaries at $\tau = \pm
\pi/4$).  What the boundary condition on these should be is not
clear.  The appearance of boundaries is not unique to compact Fermi
surfaces; boundaries of this type, both timelike and spacelike, have
appeared in the analysis of noncompact Fermi surfaces in
\cite{Alexandrov:2003uh}.

Perhaps it is possible to find a solution to the effective spacetime
theory which would mimic the properties of the droplet cosmology
outlined above.  This intriguing question is left for future
research.

\section*{Acknowledgements}We are grateful to A. Strominger for
helpful discussions. This work was supported in part by DOE grant
DE-FG02-91ER40654. The work of JML is supported by an NDSEG
Fellowship sponsored by the Department of Defense.

\section*{Appendix}
\label{sec:appendix}

In this Appendix, we will analyze an example which does not fall
into the restricted category of solutions analyzed in section
\ref{sec:static}. We return to the general case from section
\ref{sec:existence}, and, using only the property (\ref{eq:tau.pm}),
we write the cubic part of the action as \bea S_{(3)} &=&
\frac{\sqrt{\pi}}{2}\int d\sigma d\tau {1 \over 6 \varphi_0
|\partial_x \tau^+ \partial_x \tau^-|} \Big\{
 \left( (\partial_x \tau^-)^3 - (\partial_x \tau^+)^3 \right )
\left( (\partial_\sigma \eta)^3 + 3\partial_\sigma\eta
(\partial_\tau \eta)^2 \right) \nonumber \\ & & \qquad-\left(
(\partial_x \tau^+)^3 + (\partial_x \tau^-)^3 \right ) \left(
3(\partial_\sigma \eta)^2\partial_\tau \eta + (\partial_\tau \eta)^3
\right) \Big\}~. \eea For the couplings in this action to be time
independent, as in equation (\ref{eq:Sstatic}),
$\partial_x\tau^{\pm} / \varphi_0$ must be a function of $\sigma$
only. We will analyze this condition in a specific example.

Consider the Fermi surface given by \be
x^2 - p^2 = 1 +(x-p)^3 e^{3t}~. \ee Parametrically, this surface
is given by \bea x &=& \cosh \omega + \frac{1}{2} e^{3t-2\omega}
\\ \nonumber p &=& \sinh \omega + \frac{1}{2} e^{3t-2\omega}~. \eea
Since the parametric form is similar to the one given in
\cite{Alexandrov:2003uh}, we use the procedure given there to
define the Alexandrov coordinates  \be \tau^+ =
t-\omega~,~~~\tau^- = t- \tilde \omega~, \ee where $\tilde \omega$
is defined by $x(\tilde \omega, t) = x(\omega, t)$ as well as
$p(\omega, t) = p_{+}$ and $p(\tilde \omega, t) = p_{-}$.  It is
possible to solve for $x$, $t$ and $p_{\pm}$ as functions of
$\tau^\pm$: \bea x(\tau^\pm) &=& -\frac{e^{2\tau^+ + 2\tau^-} -
e^{\tau^+} - e^{\tau^-}}
{2\sqrt{e^{\tau^+ + \tau^-} - e^{2\tau^+ + 3\tau^-} - e^{3\tau^+ + 2\tau^-}}} \nonumber \\
\exp(t(\tau^\pm)) &=& -\frac
{\sqrt{e^{\tau^+ + \tau^-} - e^{2\tau^+ + 3\tau^-} - e^{3\tau^+ + 2\tau^-}}}
{e^{2\tau^+ + \tau^-} + e^{\tau^+ + 2\tau^-} - 1} \nonumber \\
p_{+}(\tau^\pm) &=& \frac{e^{2\tau^+ + 2\tau^-} +
2e^{3\tau^+ + \tau^-} +e^{\tau^-} - e^{\tau^+}}
{2\sqrt{e^{\tau^+ + \tau^-} - e^{2\tau^+ + 3\tau^-} - e^{3\tau^+ + 2\tau^-}}} \nonumber \\
p_{-}(\tau^\pm) &=& \frac{e^{2\tau^+ + 2\tau^-} +
2e^{3\tau^- + \tau^+} +e^{\tau^+} - e^{\tau^-}}
{2\sqrt{e^{\tau^+ + \tau^-} - e^{2\tau^+ + 3\tau^-} - e^{3\tau^+ + 2\tau^-}}}~.
\eea

The coordinates given here have the property that the edge of
the Fermi sea ($p_{+} = p_{-}$) is at $2 \sigma = \tau^+ - \tau^-
= 0$. It is now possible to compute $\partial_x\tau^{\pm} /
\varphi_0$. Not surprisingly, this is not a function of $\sigma$
only. The question is whether, by a suitable conformal change of
coordinates to $\bar \tau^\pm$, this condition could be satisfied.
The change of coordinates would have to map $\sigma = 0$ to
itself to maintain a static Fermi sea edge in the new
coordinates.  Thus, the change of coordinates must be of the form
$\tau^\pm = f(\bar \tau^\pm)$, with $f(\cdot)$ an arbitrary
function.  Define $Q_\pm \equiv \partial_x\tau^{\pm} / \varphi_0$. The
necessary condition is then \be 0 = \partial_{\bar \tau} Q_\pm =
f'(\bar \tau^+) \partial_{\tau^+} Q_\pm + f'(\bar \tau^-)
\partial_{\tau^-} Q_\pm \ee implying that $\partial_{\tau^-} Q_\pm /
\partial_{\tau^+} Q_\pm$ is of the form \be W_\pm(\tau^+, \tau^-)  \equiv
\frac{\partial_{\tau^-} Q_\pm }{  \partial_{\tau^+} Q_\pm} = -\frac
{f'(\bar \tau^+)}{f'(\bar \tau^-)} = - \frac
{F(\tau^+)}{F(\tau^-)}~. \ee
Therefore, \be W_\pm(\tau^+, \tau^-)
W_\pm(\tau^-, \tau^+) = 1~. \ee By explicit computation, it can be
checked that this condition is not satisfied. Therefore, there does
not exist a coordinate transformation after which $S_{(3)}$ has no
$\tau$ dependence.

\end{document}